\documentstyle[preprint,aps]{revtex}

\begin{document}


\title{Matter Collineations of Some Well Known Spacetimes}

\author{M. Sharif \thanks{Permanent Address: Department of
Mathematics, Punjab University, Quaid-e-Azam Campus Lahore-54590,
PAKISTAN, e-mail address: hasharif@yahoo.com}\\ Department of
Physics, Hanyang University, Seoul 133-791, KOREA}

\maketitle

\begin{abstract}
{\it We derive matter collineations of the Bianchi types I, II,
III, VIII and IX, and Kantowski-Sachs spacetimes. It is found that
matter collineations turn out similar to Ricci collineations with
different constraint equations. We solve the constraint equation
for a particular case and obtain three cosmological models which
represent perfect fluid dust solutions.}
\end{abstract}

{\bf PACS}: 04.20Gz, 02.40Ky

\newpage

\section{Introduction}

There has been a recent literature [1-5, and references therein]
which shows a significant interest in the study of various
symmetries. These symmetries arise in the exact solutions of
Einstein field equations given by
\begin{equation}
G_{ab}=R_{ab}-\frac 12 Rg_{ab}=\kappa T_{ab},
\end{equation}
where $G_{ab}$ represents the components of Einstein tensor,
$R_{ab}$ are the components of Ricci tensor and $T_{ab}$ are the
components of matter (or energy-momentum) tensor, $R$ is the Ricci
scalar and $\kappa$ is the gravitational constant. The geometrical
nature of a spacetime is expressed by the metric tensor through
Einstein field equations. The well known connection between
Killing vectors (KVs) and constants of the motion [6,7] has
encouraged the search for general relations between collineations
and conservation laws [2].

Curvature and the Ricci tensors are the important quantities which
play a vital role in understanding the geometric structure of
spacetime. A basic work on curvature collineations (CCs) and Ricci
collineations (RCs) has been carried out by Katzin et al. [8] and
a complete classification of CCs and RCs for spherical and plane
symmetric spacetimes has been obtained by Qadir et al. [2,3].

The energy-momentum tensor represents the matter part of the
Einstein field equations and gives the matter field symmetries.
Thus the study of matter collineations (MCs) seems more relevant
from the physical point of view. Carot et al. [9] have studied
MCs, as a symmetry property of the energy-momentum tensor $T_{ab}$
and have discussed the possible MCs for the case of a degenerate
$T_{ab}$. Hall et al. [10] have presented a discussion of RCs and
MCs in spacetime and have suggested for the evaluation of matter
symmetries to see the similarities between RCs and MCs of the
spacetime. Recently, Yavuz et al. have discussed the RCs for the
Bianchi types II, VIII and IX [11] and for the Bianchi types I and
III, and Kantowski-Sachs spacetimes [12]. In this paper, we
address the problem of calculating MCs for these spacetimes and
establish the relation between KVs, RCs and MCs. We shall present
the complete procedure for solving the MC equations for the
Bianchi types I and III, and Kantowski-Sachs spacetimes. As the
same methods apply for the Bianchi types II, VIII and IX
spacetimes, we do not solve them explicitly and only give the
results.

The break of the paper follows. In the next section we shall write
down the set of MC equations for the Bianchi types I and III, and
Kantowski-Sachs metrics and in section three we shall solve them.
In section four, we shall solve MC equations for the Bianchi types
II, VIII and IX metrics. Finally, a summary of the results
obtained will be presented.

\section{Matter Collineation Equations}

A vector $\xi$ is called a MC if the Lie derivative of the
energy-momentum tensor along that vector is zero. That is,
\begin{equation}
{\cal L}_\xi T=0,
\end{equation}
where $T$ is the energy-momentum tensor and ${\cal L_\xi}$ denotes
the Lie derivative along $\xi$ of the energy-momentum tensor $T$.
This equation, in a torsion-free space in a coordinate basis,
reduces to a simple partial differential equation (PDE),
\begin{equation}
T_{ab,c}\xi^c+T_{ac}\xi^c_{,b}+T_{bc}\xi^c_{,a}=0,\quad a,b,c=0,1,2,3.
\end{equation}
where $,$ denotes partial derivative with respect to the
respective coordinate. These are ten coupled PDEs for four unknown
functions $(\xi^a)$ which are functions of all spacetime
coordinates.

The metric for the Bianchi types I ($\delta=0$) and III
($\delta=-1$), and Kantowski-Sachs ($\delta=+1$) cosmological
models is given by [13]-[15].
\begin{equation}
ds^2=dt^2-A^2(t)dr^2-B^2(t)(d\theta^2+f^2(\theta)d\phi^2),
\end{equation}
where $f(\theta)$ is $\theta, \sinh\theta$ or $\sin\theta$
according as $\delta =-\frac{f''}{f}=0,-1$ or $+1$ respectively
and prime denotes differentiation with respect to $\theta$. Notice
that $f^2(\frac{f'}{f})'=-1\Leftrightarrow (ff')'=2f'^2-1$. The
non-vanishing components of the energy-momentum tensor for the
above metric are the following
\begin{equation}
T_{00}=2\frac{\dot{A}\dot{B}}{AB}+\frac{\dot{B}^2}{B^2}+\frac{\delta}{B^2},
\end{equation}
\begin{equation}
T_{11}=-A^2(\frac{2\ddot{B}}{B}+\frac{\dot{B}^2}{B^2}+\frac{\delta}{B^2}),
\end{equation}
\begin{equation}
T_{22}=-B^2(\frac{\ddot{A}}{A}+\frac{\ddot{B}}{B}+\frac{\dot{A}\dot{B}}{AB}),
\end{equation}
\begin{equation}
T_{33}=f^2T_{22},
\end{equation}
where dot denotes differentiation with respect to the time
coordinate $t$. It is to be noted that we have taken $\kappa=1$
for simplicity. Using Eqs.(3) and (5)-(8), the MC equations will
become
\begin{equation}
(M_{00}):\quad T_{00,0}\xi^0+2T_{00}\xi^0_{,0}=0,
\end{equation}
\begin{equation}
(M_{11}):\quad T_{11,0}\xi^0+2T_{11}\xi^1_{,1}=0,
\end{equation}
\begin{equation}
(M_{22}):\quad T_{22,0}\xi^0+2T_{22}\xi^2_{,2}=0,
\end{equation}
\begin{equation}
(M_{33}):\quad fT_{22,0}\xi^0+2f'T_{22}\xi^2+2fT_{22}\xi^3_{,3}=0,
\end{equation}
\begin{equation}
(M_{01}):\quad T_{00}\xi^0_{,1}+T_{11}\xi^1_{,0}=0,
\end{equation}
\begin{equation}
(M_{02}):\quad T_{00}\xi^0_{,2}+T_{22}\xi^2_{,0}=0,
\end{equation}
\begin{equation}
(M_{03}):\quad T_{00}\xi^0_{,3}+f^2T_{22}\xi^3_{,0}=0,
\end{equation}
\begin{equation}
(M_{12}):\quad T_{11}\xi^1_{,2}+T_{22}\xi^2_{,1}=0,
\end{equation}
\begin{equation}
(M_{13}):\quad T_{11}\xi^1_{,3}+f^2T_{22}\xi^3_{,1}=0,
\end{equation}
\begin{equation}
(M_{23}):\quad \xi^2_{,3}+f^2\xi^3_{,2}=0.
\end{equation}

\section{Solution of the MC Equations}

We solve the MC equations (9)-(18) for the following three
cases:\\ (1) One component of $\xi^a(x^b)$ is different from zero;
\\ (2) Two components of $\xi^a(x^b)$ are different from zero; \\
(3) Three components of $\xi^a(x^b)$ are different from zero.

\subsection{One Component of $\xi^a(x^b)$ is Different from
Zero}

This case has the following four possibilities: \\ (1a)
$\xi^a=(\xi^0(x^b),0,0,0)$;\\ (1b) $\xi^a=(0,\xi^1(x^b),0,0)$;\\
(1c) $\xi^a=(0,0,\xi^2(x^b),0)$;\\ (1d)
$\xi^a=(0,0,0,\xi^3(x^b))$.

In the case (1a), we have $\xi^0=\frac{c_0}{\sqrt{|T_{00}|}}$,
where $c_0$ is an arbitrary constant. Using Eqs.(10) and (11), it
follows that $T_{11}=constant=T_{22}$. Thus, in this case, we have
one MC, i.e.,
$\xi_{(1)}=\frac{1}{\sqrt{|T_{00}|}}\frac{\partial}{\partial t}$.
In the cases (1b) and (1d), $\xi^1$ and $ \xi^3$ respectively
become $constants$. For the case (1c), using Eq.(12) we have
$f'T_{22}=0$ which implies that either $f$ is constant or
$T_{22}=0$. If $f$ is constant then $\xi^2$ also becomes
$constant$ and if $T_{22}=0$ then $\xi^2$ becomes an arbitrary
function of $x^a$. It follows that in each case (1a-d), MC will be
one.

\subsection{Two Components of $\xi^a(x^b)$ are Different from Zero}

In this case, we have six different possibilities:\\ (2a)
$\xi^a=(\xi^0(x^b),\xi^1(x^b),0,0)$;\\ (2b)
$\xi^a=(\xi^0(x^b),0,\xi^2(x^b),0)$;\\ (2c)
$\xi^a=(\xi^0(x^b),0,0,\xi^3(x^b))$;\\ (2d)
$\xi^a=(0,\xi^1(x^b),\xi^2(x^b),0)$;\\ (2e)
$\xi^a=(0,\xi^1(x^b),0,\xi^3(x^b))$;\\ (2f)
$\xi^a=(0,0,\xi^2(x^b),\xi^3(x^b)).$\\ {\bf Case (2a)}
$\xi^a=(\xi^0(x^b),\xi^1(x^b),0,0)$:

In this case, from Eqs.(14)-(17) we see that $\xi^0$ and $\xi^1$
become functions of $t,r$ and from Eq.(9) we have
$\xi^0=\frac{A(r)}{\sqrt{|T_{00}|}}$, where $A(r)$ is an
integration function. Using Eq.(11) or (12), we obtain
$T_{22,0}\xi^0=0$ which implies that either (i) $T_{22,0}=0$ or
(ii) $\xi^0=0$. In the subcase (i), after replacing the value of
$\xi^0$ in Eq.(10), it follows that
\begin{equation}
\xi^1(t,r)=-\frac{T_{11,0}}{2T_{11}\sqrt{|T_{00}|}}\int
A(r)dr+B(t),
\end{equation}
where $B(t)$ is an integration function. Substituting this value
of $\xi^1$ together with $\xi^0$ in Eq.(13), we obtain
\begin{equation}
\frac{A_{,11}}{A} =\frac{T_{11}}{\sqrt{|T_{00}|}}
(\frac{T_{11,0}}{2T_{11}\sqrt{|T_{00}|}}\dot{)}=-\alpha^2,
\end{equation}
where $\alpha^2$ is a separation constant which may be positive,
negative or zero. \\ When $\alpha^2>0$, we have
\begin{eqnarray}
\xi^0=\frac{1}{\sqrt{|T_{00}|}}(c_0\cos\alpha r+c_1\sin\alpha
r),\\ \nonumber \xi^1=-\frac{T_{11,0}}{2\alpha
T_{11}\sqrt{|T_{00}|}}(c_0\sin\alpha r-c_1\cos\alpha r)+c_2,
\end{eqnarray}
where $c_0,c_1,c_2$ are arbitrary constants. It follows that MCs
can be written as
\begin{eqnarray}
\xi_{(1)}=\frac{1}{\sqrt{|T_{00}|}}\cos\alpha
r\frac{\partial}{\partial t}-\frac{T_{11,0}}{2\alpha
T_{11}\sqrt{|T_{00}|}}\sin\alpha r\frac{\partial}{\partial
r},\\\nonumber \xi_{(2)}=\frac{1}{\sqrt{|T_{00}|}}\sin\alpha
r\frac{\partial}{\partial t}+\frac{T_{11,0}}{2\alpha
T_{11}\sqrt{|T_{00}|}}\cos\alpha r\frac{\partial}{\partial
r},\quad \xi_{(3)}=\frac{\partial}{\partial r}.
\end{eqnarray}
When $\alpha^2<0$, $\alpha^2$ is replaced by $-\alpha^2$ in
Eq.(20) and we obtain the following solution
\begin{eqnarray}
\xi^0=\frac{1}{\sqrt{|T_{00}|}}(c_0\cosh\alpha r+c_1\sinh\alpha
r),\\ \nonumber \xi^1=-\frac{T_{11,0}}{2\alpha
T_{11}\sqrt{|T_{00}|}}(c_0\sinh\alpha r+c_1\cosh\alpha r)+c_2.
\end{eqnarray}
For $\alpha^2=0$, we obtain
$\frac{T_{11,0}}{2T_{11}\sqrt{|T_{00}|}}=\beta$, where $\beta$ is
an arbitrary constant. This implies that either $\beta$ is
non-zero or zero. For $\beta\neq 0$, it follows that
\begin{equation}
\xi^0=\frac{1}{\sqrt{|T_{00}|}}(c_0r+c_1),\quad
\xi^1=-\beta(c_0\frac{r^2}{2}+c_1r)+c_0\int\frac{\sqrt{|T_{00}|}}{T_{11}}dt+c_2.
\end{equation}
For $\beta=0$, we have
\begin{equation}
\xi^0=\frac{1}{\sqrt{|T_{00}|}}(c_0r+c_1),\quad
\xi^1=\frac{c_0}{T_{11}}\int \sqrt{|T_{00}|}dt+c_2, \quad
T_{11}=constant.
\end{equation}
Thus, in the subcase (2ai), MCs turn out to be three in all the
possibilities. If Eq.(20) is not satisfied by $T_{00}$ and
$T_{11}$ then $A=0$ and this reduces to the case (1b). For the
subcase (ii) when $\xi^0=0$, it also reduces to the case (1b).\\
{\bf Case (2b)} $\xi^a=(\xi^0(x^b),0,\xi^2(x^b),0)$:

In this case, we see from Eqs.(13),(15),(16) and (18) that $\xi^0$
and $\xi^2$ become functions of $t,\theta$ and from Eq.(9) we have
$\xi^0=\frac{A(\theta)}{\sqrt{|T_{00}|}}$, where $A(\theta)$ is an
integration function. It follows from Eq.(10) that either (i)
$T_{11,0}=0$ or (ii) $\xi^0=0$. In the subcase (i), subtracting
Eqs.(11) and (12), we have
\begin{equation}
\xi^2(t,\theta)=fB(t),
\end{equation}
where $B(t)$ is an integration function. Plugging the values of
$\xi^0$ and $\xi^2$ in Eq.(11) we obtain
\begin{equation}
\frac{T_{22,0}}{2T_{22}\sqrt{|T_{00}|}}\frac{1}{B}=-\frac{f'}{A}=\alpha^2,
\end{equation}
where $\alpha^2$ is a separation constant which is not zero. We
can choose $\alpha^2=1$ without loss of generality then the values
of $A$ and $B$ can be found from Eq.(27). Using these values of
$A$ and $B$, $\xi^0$ and $\xi^1$ will take the form
\begin{equation}
\xi^0=-\frac{f'}{\sqrt{|T_{00}|}},\quad
\xi^2=\frac{fT_{22,0}}{2T_{22}\sqrt{|T_{00}|}}.
\end{equation}
If we make use of Eq.(28) in Eq.(14), we obtain
\begin{equation}
\frac{T_{22}}{\sqrt{|T_{00}|}}(\frac{T_{22,0}}{2T_{22}\sqrt{|T_{00}|}}\dot{)}
=\frac{f''}{f}=-\delta,
\end{equation}
where $\delta$ can take values of $0,-1$ or $+1$ according as
Bianchi types I, III or Kantowski-Sachs metrics respectively. For
$\delta=0$, we have
$\frac{T_{22,0}}{2T_{22}\sqrt{|T_{00}|}}=\beta$, where $\beta$ is
an integration constant. This gives rise to two different
possibilities either $\beta\neq 0$ or $\beta=0$. In the first
subcase we have
\begin{equation}
\xi^0=\frac{c_0}{\sqrt{|T_{00}|}},\quad \xi^2=c_1\beta\theta.
\end{equation}
Thus the MCs will be two which are
$\xi_{(1)}=\frac{1}{\sqrt{|T_{00}|}}\frac{\partial}{\partial
t},\quad\xi_{(2)}=\beta\theta\frac{\partial}{\partial \theta}$.
For the second subcase, we have the results of (1a). The subcase
(ii), when $\xi^0=0$, reduces to the case (1b).\\ {\bf Case (2c)}
$\xi^a=(\xi^0(x^b),0,0,\xi^3(x^b))$:

In this case, using Eqs.(13),(14),(17) and (18), it turns out that
$\xi^0$ and $\xi^3$ are functions of $t,\phi$ and from Eq.(9), it
follows that $\xi^0=\frac{A(\phi)}{\sqrt{|T_{00}|}}$, where
$A(\phi)$ is an integration function. From Eq.(10) or (11) we see
that either $T_{11,0}=0=T_{22,0}$ or $\xi^0=0$. In the first
subcase , we have
\begin{equation}
\xi^0=\frac{c_0}{\sqrt{|T_{00}|}},\quad \xi^3=c_1.
\end{equation}
It follows that the MCs turn out to be two, i.e.,
$\xi_{(1)}=\frac{1}{\sqrt{|T_{00}|}}\frac{\partial}{\partial
t},\quad\xi_{(2)}=\frac{\partial}{\partial \phi}$. The second
subcase reduces to (1c).\\ {\bf Case (2d)}
$\xi^a=(0,\xi^1(x^b),\xi^2(x^b),0)$:

In this case, we see from Eqs.(10), (11), (13), (14), (17) and
(18) that $\xi^1$ and $\xi^2$ become functions of $\theta$ and $r$
respectively. Eq.(12) implies that either $T_{22}=0$ or $\xi^2=0$.
If $T_{22}=0$, we have $ \xi^1=c_0,\quad \xi^2=\xi(x^a)$. Thus MCs
are $\xi=\frac{\partial}{\partial
r}+\xi^2(x^a)\frac{\partial}{\partial \theta}$ which shows that
MCs are arbitrary in the $\theta$ direction. When $T_{11}=0$ then
both $\xi^1$ and $\xi^2$ become arbitrary functions of four vector
$x^a$. It follows that $\xi=\xi^1(x^a)\frac{\partial}{\partial
r}+\xi^2(x^a)\frac{\partial}{\partial \theta}$ which gives
arbitrary MCs in both radial and $\theta$ directions. In the
second subcase, when $\xi^2=0$, it reduces to (1b) if
$T_{11}\neq0$ and for $T_{11}=0$, $\xi^1$ become arbitrary
function of $x^a$ which gives
$\xi=\xi^1(x^a)\frac{\partial}{\partial r}$.

When $T_{11}=0=T_{22}$, using Eqs.(6) and (7), it follows that
\begin{equation}
\frac{2\ddot{B}}{B}+\frac{\dot{B}^2}{B^2}+\frac{\delta}{B^2}=0.
\end{equation}
\begin{equation}
\frac{\ddot{A}}{A}+\frac{\ddot{B}}{B}+\frac{\dot{A}\dot{B}}{AB}=0.
\end{equation}
Plugging the value of $\frac{\ddot{B}}{B}$ from Eq.(33) in
Eq.(32), we obtain
\begin{equation}
2\frac{\ddot{A}}{A}+2\frac{\dot{A}\dot{B}}{AB}-\frac{\dot{B}^2}{B^2}-\frac{\delta}{B^2}=0.
\end{equation}
This is a second order non-linear differential equation in $A$ and
$B$ and can only be solved by assuming some relation between these
two functions. If we choose $B=cA$, where $c$ is an arbitrary
constant, then Eq.(34) will become
\begin{equation}
2A\ddot{A}+\dot{A}^2=\frac{\delta}{c^2}.
\end{equation}
Now assume that $A=(at+b)^m$, where $a,b$ and $m$ are arbitrary
constants, then using Eq.(35), it follows that only possible
solutions for $\delta=0,+1$ and $-1$ are $m=\frac
23,m=1(a^2c^2=1)$ and $m=1(a^2c^2=-1)$ respectively. Thus for the
Bianchi type I $(\delta=0)$ metric, we have
\begin{equation}
ds^2=dt^2-(at+b)^{\frac 23}dr^2-c(at+b)^{\frac 23}
(d\theta^2+\theta^2d\phi^2).
\end{equation}
For the Bianchi type III $(\delta=-1)$ metric, it follows that
\begin{equation}
ds^2=dt^2-(at+b)dr^2-c(at+b)(d\theta^2+\sinh^2\theta d\phi^2),
\end{equation}
where $a^2c^2=-1$. For the Kantowski-Sachs $(\delta=+1)$ metric,
we obtain
\begin{equation}
ds^2=dt^2-(at+b)dr^2-c(at+b)(d\theta^2+\sin^2\theta d\phi^2),
\end{equation}
where $a^2c^2=+1$. It can easily be verified that all these
spacetimes represent perfect fluid dust solutions. The energy
density for each of the above metrics are given as
\begin{equation}
\rho_0=\frac{4a^2}{3(at+b)^2},\quad
\rho_{-1}=\frac{4+3(at+b)^{\frac 23}}{3(at+b)^2},\quad
\rho_{+1}=-\frac{4+3(at+b)^{\frac 23}}{3(at+b)^2}.
\end{equation}
It is interesting to note that the energy density is positive for
the Bianchi types I and III metrics.\\ {\bf Case (2e)}
$\xi^a=(0,\xi^1(x^b),0,\xi^3(x^b))$:

It follows from Eqs.(10),(12),(13),(15),(16) and (18) that $\xi^1$
and $\xi^3$ are functions of $\phi$ and $r$ respectively. Using
Eq.(17), it turns out that $\xi^1$ and $\xi^3$ become $constants$.
Thus the two MCs will be $\xi_{(1)}=\frac{\partial}{\partial
r},\quad \xi_{(2)}=\frac{\partial}{\partial \phi}$.
\\ {\bf Case (2f)} $\xi^a=(0,0,\xi^2(x^b),\xi^3(x^b))$:

In this case, using Eqs.(11), (12), (14)-(17) we find that there
arise two possible situations either $T_{22}=0$ or $T_{22}\neq 0$.
For the first option, $\xi^2$ and $\xi^3$ become arbitrary
functions of $x^a$. For the second option, it follows from
Eqs.(11), (14)-(17) that $\xi^2=\xi^2(\phi)$ and
$\xi^3=\xi^3(\theta,\phi)$. Using Eqs.(12) and (18), it can be
shown, after some algebra, that
\begin{equation}
\xi^2=c_0\cos\phi+c_1\sin\phi,\quad
\xi^3=-\frac{f'}{f}(c_0\sin\phi-c_1\cos\phi)+c_2.
\end{equation}
Thus the MCs are given as
\begin{equation}
\xi=c_0(\cos\phi\frac{\partial}{\partial\theta}-\frac{f'}{f}\sin\phi
\frac{\partial}{\partial\phi})+c_1(\sin\phi\frac{\partial}{\partial\theta}
+\frac{f'}{f}\cos\phi)\frac{\partial}{\partial\phi}+c_2\frac{\partial}{\partial\phi}.
\end{equation}
It follows that the MCs are three given by
\begin{equation}
\xi_{(1)}=\cos\phi\frac{\partial}{\partial\theta}-\frac{f'}{f}\sin\phi
\frac{\partial}{\partial\phi},\quad
\xi_{(2)}=\sin\phi\frac{\partial}{\partial\theta}
+\frac{f'}{f}\cos\phi\frac{\partial}{\partial\phi},\quad
\xi_{(3)}=\frac{\partial}{\partial\phi}.
\end{equation}
which also represent the generators of a group $G_3$. These are
just the KVs associated with spherical symmetry of the Bianchi
types I and III, and Kantowski-Sachs spacetimes for
$\delta=0,-1,+1$ respectively [13].

\subsection{Three Components of $\xi^a(x^b)$ are Different from Zero}

It has four different possibilities:\\ (3a)
$\xi^a=(\xi^0(x^b),\xi^1(x^b),\xi^2(x^b),0)$;\\ (3b)
$\xi^a=(\xi^0(x^b),\xi^1(x^b),0,\xi^3(x^b))$;\\ (3c)
$\xi^a=(\xi^0(x^b),0,\xi^2(x^b),\xi^3(x^b))$;\\ (3d)
$\xi^a=(0,\xi^1(x^b),\xi^2(x^b),\xi^3(x^b))$;\\ {\bf Case (3a)}
$\xi^a=(\xi^0(x^b),\xi^1(x^b),\xi^2(x^b),0)$:

In this case, using Eqs.(15), (17) and (18) we find that
$\xi^0,\xi^1$ and $\xi^2$ are functions of $t,r,\theta$ and from
Eq.(9) we have $\xi^0=\frac{A(r,\theta)}{\sqrt{|T_{00}|}}$, where
$A(r,\theta)$ is an integration function. If we make use of this
value of $\xi^0$ in Eq.(12), we obtain
$\xi^2=-\frac{T_{22,0}}{2T_{22}\sqrt{|T_{00}|}}\frac{Af}{f'}$.
Substituting this value of $\xi^2$ together with $\xi^0$ in
Eq.(11), we obtain the value of $A$ as follows
\begin{equation}
A(r,\theta)=A_1(r)f',
\end{equation}
where $A_1(r)$ is an integration function. From Eq.(10), it
follows that
\begin{equation}
\xi^1=-\frac{T_{11,0}}{2T_{11}\sqrt{|T_{00}|}}f'\int
A_1dr+B(t,\theta),
\end{equation}
where $B(t,\theta)$ is an integration function. Using values of
$\xi^0$ and $\xi^2$ in Eq.(14), we have
\begin{equation}
\frac{T_{22}}{\sqrt{|T_{00}|}}(\frac{T_{22,0}}{2T_{22}\sqrt{|T_{00}|}}\dot{)}
=\frac{f''}{f}=-\delta.
\end{equation}
Now plugging the values of $\xi^0$ and $\xi^1$ in Eq.(13), we
obtain
\begin{equation}
\frac{T_{11}}{\sqrt{|T_{00}|}}(\frac{T_{11,0}}{2T_{11}{\sqrt{|T_{00}|}}}\dot{)}
=\frac{A_{1,11}}{A_1}=-\alpha^2,
\end{equation}
where $\alpha^2$ is a separation constant which may be positive,
negative or zero. Thus there arise six different possibilities.
(i) $\alpha^2>0,\delta\neq 0$;\quad(ii) $\alpha^2<0,\delta\neq
0$;\quad(iii) $\alpha^2=0,\delta\neq 0$;\quad(iv)
$\alpha^2>0,\delta=0$;\quad(v) $\alpha^2<0,\delta=0$;\quad(vi)
$\alpha^2=0,\delta=0$.\\ For the subcase (i), after some algebraic
manipulation, it is shown that
\begin{eqnarray}
\xi^0=\frac{1}{\sqrt{|T_{00}|}}(c_0\cos\alpha r+c_1\sin\alpha
r)f',\\\nonumber \xi^1=-\frac{f'}{\alpha}\frac{T_{11,0}}{2
T_{11}\sqrt{|T_{00}|}}(c_0\sin\alpha r-c_1\cos\alpha
r)+c_2,\\\nonumber \xi^2=-\frac{T_{22,0}}{2
T_{22}\sqrt{|T_{00}|}}(c_0\cos\alpha r+c_1\sin\alpha r)f,\quad
T_{22}=\frac{\delta}{\alpha^2}T_{11}+c,
\end{eqnarray}
where $c$ is an arbitrary constants and $\delta$ can take values
$\pm1$. It follows that MCs are three which can be written as
\begin{eqnarray}
\xi_{(1)}=\frac{f'}{\sqrt{|T_{00}|}}\cos\alpha
r\frac{\partial}{\partial t} -\frac{f'}{\alpha}\frac{T_{11,0}}{2
T_{11}\sqrt{|T_{00}|}}\sin\alpha r\frac{\partial}{\partial
r}-\frac{fT_{22,0}}{2 T_{22}\sqrt{|T_{00}|}}\cos\alpha
r\frac{\partial}{\partial \theta},\\\nonumber
\xi_{(2)}=\frac{f'}{\sqrt{|T_{00}|}}\sin\alpha
r\frac{\partial}{\partial t} +\frac{f'}{\alpha}\frac{T_{11,0}}{2
T_{11}\sqrt{|T_{00}|}}\cos\alpha r\frac{\partial}{\partial
r}-\frac{fT_{22,0}}{2 T_{22}\sqrt{|T_{00}|}}\sin\alpha
r\frac{\partial}{\partial \theta},\quad
\xi_{(3)}=\frac{\partial}{\partial r}.
\end{eqnarray}
In the subcase (ii), $\alpha^2$ is replaced by $-\alpha^2$ in
Eq.(38) and it follows that
\begin{eqnarray}
\xi^0=\frac{1}{\sqrt{|T_{00}|}}(c_0\cosh\alpha r+c_1\sinh\alpha
r)f',\\\nonumber \xi^1=-\frac{f'}{\alpha}\frac{T_{11,0}}{2
T_{11}\sqrt{|T_{00}|}}(c_0\sinh\alpha r+c_1\cosh\alpha
r)+c_2,\\\nonumber \xi^2=-\frac{T_{22,0}}{2
T_{22}\sqrt{|T_{00}|}}(c_0\cosh\alpha r+c_1\sinh\alpha r)f,\quad
T_{22}=\frac{\delta}{\alpha^2}T_{11}+c.
\end{eqnarray}
For the third subcase, we have $A_1=c_0r+c_1$ and
$\frac{T_{11,0}}{2T_{11}\sqrt{|T_{00}|}}=\beta$, where $\beta$ is
an arbitrary constant. This implies that either $\beta$ is
non-zero or zero. If $\beta\neq 0$, it reduces to the case (1b).
For $\beta=0$, we have $T_{11}=constant$. Using these values in
Eq.(13) and (16) we obtain
\begin{eqnarray}
\xi^0=\frac{1}{\sqrt{|T_{00}|}}(c_0r+c_1)f',\quad
\xi^1=-\frac{c_0f'T_{22,0}}{2\delta T_{11}
\sqrt{|T_{00}|}}+c_2,\quad \xi^2=-\frac{T_{22,0}f}{2T_{22}
\sqrt{|T_{00}|}}(c_0r+c_1),\\\nonumber T_{11}=constant,\quad
T_{22,0}=2\delta\sqrt{|T_{00}|}\int\sqrt{|T_{00}|}dt.
\end{eqnarray}
The fourth and fifth subcases reduce to (1b).

In the subcase (vi), we obtain $A_1=c_0r+c_1,\quad
\frac{T_{11,0}}{2T_{11}\sqrt{|T_{00}|}}=\beta$ and $
\frac{T_{22,0}}{2T_{22}\sqrt{|T_{00}|}}=\gamma$, where $\beta$ and
$\gamma$ are arbitrary constants. We see from Eq.(16) that $\gamma
T_{22}=0$ which implies that either $\gamma=0$ or $T_{22}=0$. This
gives rise to three possibilities that $(\star)\quad\gamma\neq
0\neq\beta, T_{22}=0$, $(\dagger)\quad\gamma=0,\beta\neq 0\neq
T_{22}$ and $(\ddagger)\quad\gamma=0=\beta, T_{22}\neq 0$. In the
subcase (vi$\star$), for the Bianchi type I spacetime, we have
\begin{equation}
\xi^0=\frac{1}{\sqrt{|T_{00}|}}(c_0r+c_1),\quad
\xi^1=-\beta(c_0\frac{r^2}{2}+c_1r)-c_0\int\frac{\sqrt{|T_{00}|}}{T_{11}}dt+c_2,\quad
\xi^2=-\gamma\theta(c_0r+c_1).
\end{equation}
For the Bianchi type II and Kantowski-Sachs spacetimes $c_0=0$.
The subcases (vi$\dagger$) and (vi$\ddagger$) reduce to the case
(2a). In the former subcase $T_{11}$ becomes $constant$ while in
the latter subcase both $T_{11}$ and $T_{22}$ become
$constants$.\\ {\bf Case (3b)}
$\xi^a=(\xi^0(x^b),\xi^1(x^b),0,\xi^3(x^b))$:

In this case, using Eqs.(14), (16) and (18), it follows that
$\xi^0,\xi^1$ and $\xi^3$ are functions of $t,r,\phi$. Using
Eqs.(15) and (17), it can be seen that $\xi^0$ and $\xi^1$ become
functions of $t,r$ while $\xi^3$ becomes only a function of
$\phi$. Further from Eq.(12), it follows that $\xi^3$ becomes
$constant$ and $T_{22,0}\xi^0=0$ which implies that either
$T_{22,0}=0$ or $\xi^0=0$. If $\xi^0=0$, this reduces to the case
(2e). However, for $\xi^0\neq 0$ $(T_{22,0}=0)$, it follows that
it gives the results of (2a) together with $\xi^3=constant$.
\\ {\bf Case (3c)} $\xi^a=(\xi^0(x^b),0,\xi^2(x^b),\xi^3(x^b))$:

In this case, we see from Eqs.(13), (16) and (17) that
$\xi^0,\xi^2$ and $\xi^3$ are functions of $t,\theta,\phi$.
Further from Eq.(10) we have either $T_{11,0}=0$ or $\xi^0=0$.
After some algebra, it can be shown that both cases reduce to the
case (2f).\\ {\bf Case (3d)}
$\xi^a=(0,\xi^1(x^b),\xi^2(x^b),\xi^3(x^b))$:

In this case, using Eqs.(10), (11), (13)-(15) we find that
$\xi^1,\xi^2$ and $\xi^3$ are functions of
$(\theta,\phi),(r,\phi)$ and $r,\theta,\phi$ respectively. Eq.(16)
gives
\begin{equation}
\xi^1=A_1(\phi)\theta+A_2(\phi),\quad
\xi^2=-arA_1(\phi)+A_3(\phi),\quad T_{11}=aT_{22},
\end{equation}
where $A_1(\phi)$ is an integration function and $a$ is an
arbitrary constant. Using Eq.(17), it follows that
\begin{equation}
\xi^3=-\frac{ar}{f}(A_{,3}\theta+A_{2,3})+B_1(\theta,\phi),
\end{equation}
where $B_1(\theta,\phi)$ is an integration function. If $a=0$,
then $T_{11}=0$ and it follows that $\xi^1$ is an arbitrary
function of four vector $x^a$ and $\xi^2, \xi^3$ will remain the
same as in the case (2f).\\ If $a\neq 0$ and $\delta=0$, for the
Bianchi type I metric, we have from Eq.(12)
\begin{equation}
A_1=c_0\cos\phi+c_1\sin\phi,\quad A_2=c_2\phi+c_3,\quad
B_1=-\frac{1}{\theta}\int A_3d\phi+A_4(\theta),
\end{equation}
where $A_4(\theta)$ is an integration function. Using Eq.(18), we
have
\begin{eqnarray}
\xi^1=(c_0\cos\phi+c_1\sin\phi)\theta+c_2,\\\nonumber
\xi^2=-ar(c_0\cos\phi+c_1\sin\phi)+c_3\cos\phi+c_4\sin\phi,\\\nonumber
\xi^3=\frac{ar}{\theta}(c_0\sin\phi-c_1\cos\phi)
-\frac{1}{\theta}(c_3\sin\phi-c_4\cos\phi)+c_5,
\end{eqnarray}
where $c_i(i=0,1,2,3,4,5)$ are arbitrary constants. Thus the Lie
algebra of MCs are spanned by
\begin{eqnarray}
\xi_{(1)}=\cos\phi\frac{\partial}{\partial\theta}
-\frac{1}{\theta}\sin\phi\frac{\partial}{\partial\phi},\quad
\xi_{(2)}=\sin\phi\frac{\partial}{\partial\theta}
+\frac{1}{\theta}\cos\phi\frac{\partial}{\partial\phi},\quad
\xi_{(3)}=\frac{\partial}{\partial\phi},\\\nonumber
\xi_{(4)}=\theta\cos\phi\frac{\partial}{\partial
r}-ar\xi_{(1)},\quad
\xi_{(5)}=\theta\sin\phi\frac{\partial}{\partial
r}-ar\xi_{(2)},\quad \xi_{(6)}=\frac{\partial}{\partial r}.
\end{eqnarray}
Notice that MCs $\xi_{(1)},\xi_{(2)}$ and $\xi_{(3)}$ correspond
to KVs associated with spherical symmetry of the Bianchi type I
spacetime while $\xi_{(4)},\xi_{(5)}$ and $\xi_{(6)}$ are the
proper MCs of the Bianchi type I spacetime. The non-vanishing
commutators are given as
\begin{eqnarray}
[\xi_{(1)},\xi_{(3)}]=\xi_{(2)},\quad
[\xi_{(2)},\xi_{(3)}]=-\xi_{(1)},\quad
[\xi_{(1)},\xi_{(4)}]=[\xi_{(2)},\xi_{(5)}]=\xi_{(6)},\\\nonumber
[\xi_{(1)},\xi_{(6)}]=[\xi_{(3)},\xi_{(5)}]=\xi_{(4)},\quad
[\xi_{(3)},\xi_{(4)}]=-\xi_{(5)},\quad
[\xi_{(4)},\xi_{(6)}]=a\xi_{(1)},\\\nonumber
[\xi_{(4)},\xi_{(5)}]=-a\xi_{(3)},\quad
[\xi_{(5)},\xi_{(6)}]=a\xi_{(2)}.\\\nonumber
\end{eqnarray}
If $a\neq 0\neq\delta$, for the Bianchi type III and
Kantowski-Sachs spacetimes, then from Eq.(12) we have $A_1=0,
A_2=c_0\phi+c_1$ and $B_1=-\frac{f'}{f}\int A_3d\phi+A_4(\theta)$.
Using Eq.(18), it follows that
\begin{equation}
\xi^1=c_0, \quad \xi^2=c_1\cos\phi+c_2\sin\phi,\quad \xi^3=
-\frac{f'}{f}(c_1\sin\phi-c_2\cos\phi)+c_3.
\end{equation}
The MCs will, therefore, become
\begin{equation}
\xi_{(1)}=\cos\phi\frac{\partial}{\partial\theta}
-\frac{f'}{f}\sin\phi\frac{\partial}{\partial\phi},\quad
\xi_{(2)}=\sin\phi\frac{\partial}{\partial\theta}
+\frac{f'}{f}\cos\phi\frac{\partial}{\partial\phi},\quad
\xi_{(3)}=\frac{\partial}{\partial\phi},\quad
\xi_{(4)}=\frac{\partial}{\partial r}.
\end{equation}
which are also the generators of a group $G_4$. We see that
$\xi_{(1)},\xi_{(2)}$ and $\xi_{(3)}$ are the same as in (2f).
Thus the MCs $\xi_{(1)},\xi_{(2)},\xi_{(3)}$ and $\xi_{(4)}$ are
non-proper.

\section{MCs of the Bianchi types II, VIII, IX Spacetimes}

The locally rotationally symmetric metric for the spatially
homogeneous Bianchi types II ($\delta=0$), VIII ($\delta=-1$) and
IX ($\delta=+1$) cosmological models can generally be written in
the form [16,17]
\begin{equation}
ds^2=dt^2-S^2(t)(dx-h(y)dz)^2-R^2(t)(dy^2+f^2(y)dz^2),
\end{equation}
where $f(y)$ is $y, \sin y$ or $\sinh y$ according as $\delta
=-\frac{f''}{f}=0,+1$ or $-1$ respectively and $h(y)$ is
$-\frac{y}{2},\cos y$ or $-\cosh y$ for $\delta= 0,+1$ or $-1$
respectively. Here prime denotes differentiation with respect to
$y$. The non-vanishing components of energy-momentum tensor are
given as follows
\begin{equation}
T_{00}=2\frac{\dot{R}\dot{S}}{RS}+\frac{\dot{R}^2}{R^2}-\frac{S^2}{4R^4}-\frac{\delta}{R^2},
\end{equation}
\begin{equation}
T_{11}=-2\frac{\ddot{R}S^2}{R}-\frac{\dot{R}^2S^2}{R^2}+3\frac{S^4}{4R^4}
-\delta\frac{S^2}{R^2},
\end{equation}
\begin{equation}
T_{22}=-R\ddot{R}-\frac{R^2\ddot{S}}{S}-\frac{R\dot{R}\dot{S}}{S}-\frac{S^2}{4R^2},
\end{equation}
\begin{equation}
T_{33}=h^2T_{11}+f^2T_{22},
\end{equation}
\begin{equation}
T_{13}=-hT_{11}.
\end{equation}
The MC equations will become
\begin{equation}
(M_{00}):\quad T_{00,0}\xi^0+2T_{00}\xi^0_{,0}=0,
\end{equation}
\begin{equation}
(M_{11}):\quad
T_{11,0}\xi^0+2(T_{11}\xi^1_{,1}-hT_{11}\xi^3_{,1})=0,
\end{equation}
\begin{equation}(M_{22}):\quad T_{22,0}\xi^0+2T_{22}\xi^2_{,2}=0,
\end{equation}
\begin{equation}
(M_{33}):\quad
T_{33,0}\xi^0+T_{33,2}\xi^2+2(-hT_{11}\xi^1_{,3}+T_{33}\xi^3_{,3})=0,
\end{equation}
\begin{equation}
(M_{01}):\quad
T_{00}\xi^0_{,1}+T_{11}\xi^1_{,0}-hT_{11}\xi^3_{,0}=0,
\end{equation}
\begin{equation}
(M_{02}):\quad T_{00}\xi^0_{,2}+T_{22}\xi^2_{,0}=0,
\end{equation}
\begin{equation}
(M_{03}):\quad
T_{00}\xi^0_{,3}-hT_{11}\xi^1_{,0}+T_{33}\xi^3_{,0}=0,
\end{equation}
\begin{equation}
(M_{12}):\quad
T_{11}\xi^1_{,2}-hT_{11}\xi^3_{,2}+T_{22}\xi^2_{,1}=0,
\end{equation}
\begin{equation}
(M_{13}):\quad
hT_{11,0}\xi^0+h'T_{11}\xi^2-T_{11}\xi^1_{,3}+hT_{11}\xi^3_{,3}+hT_{11}\xi^1_{,1}
-T_{33}\xi^3_{,1}=0,
\end{equation}
\begin{equation}
(M_{23}):\quad
T_{22}\xi^2_{,3}-hT_{11}\xi^1_{,2}+T_{33}\xi^3_{,2}=0.
\end{equation}

We solve the above equations for the cases (1), (2), (3b) and (3d)
given in the last section.

In all the possibilities of the case (1), we have the similar
results as in the previous section for the Bianchi types I, III
and Kantowski-Sachs spacetimes.

In the case (2a), using Eq.(68), it follows that either
$T_{22,0}=0$ or $\xi^0=0$. For $T_{22,0}=0$, after some algebra,
we obtain $\xi^0=\frac{c_0}{\sqrt{|T_{00}|}}$, $\xi^1=c_1$ and
$T_{11}=constant$. The subcase $\xi^0=0$ reduces to the case (1b).
In the case (2b), using Eq.(67), we see that either $T_{11,0}=0$
or $\xi^0=0$. When we take $T_{11,0}=0$, it gives $\xi^2=0$ and so
reduces to the case (1a) while $\xi^0=0$ reduces to (1c). In the
case (2c), Eq.(68) yields that either $T_{22,0}=0$ or $\xi^0=0$.
If $T_{22,0}=0$, after some algebraic manipulation, it follows
that $\xi^0=\frac{c_0}{\sqrt{|T_{00}|}}$, $\xi^3=c_1$ and
$T_{11}=constant$ while the subcase $\xi^0=0$ reduces to the case
(1d). In the case (2d), we employ Eqs.(69) and (74) which show
that either $T_{11}=0=T_{22}$ or $\xi^2=0$. For the former
subcase, we obtain $\xi^1$ and $\xi^2$ as an arbitrary functions
of four vector $x^a$ while for the latter subcase, we have the
result of (1b). In the case (2e), after some algebra, we obtain
$\xi^1=c_0$, $\xi^3=c_1$. The case (2f) gives the similar results
as the case (1d).

In the case (3b), it follows from Eq.(68) that either $T_{22,0}=0$
or $\xi^0=0$. For $T_{22,0}=0$, we obtain
$\xi^0=\frac{c_0}{\sqrt{|T_{00}|}}$, $\xi^1=c_1$, $\xi^3=c_2$ and
$T_{11}=constant$ while the subcase $\xi^0=0$ reduces to the case
(2e). In the case (3d), if we choose $T_{11}=aT_{22}$, where $a$
is an arbitrary constant then we have two possibilities either
$a=0$ or $a\neq 0$. For $a=0$, after some algebraic manipulation,
we obtain
\begin{equation}
\xi^1=\xi^1(y,z),\quad \xi^2=c_0\cos z+c_1\sin z,\quad
\xi^3=-\frac{f'}{f}(c_0\sin z-c_1\cos z)+c_2
\end{equation}
and for $a\neq 0$, this reduces to the case (2e).

\section{Conclusion}

We have evaluated the MCs for the Bianchi types I, II, III, VIII
and IX, and Kantowski-Sachs spacetimes. First we discuss the MCs
for the Bianchi types I and III, and Kantowski-Sachs spacetimes.
We see that in the cases (1b-d), the MCs are identical to the KVs.
Further we note that in the case (1b), the MC represents a
translation along the radial direction $r$ while in the cases (1c)
and (1d), the MC represents rotations in $\theta$ and $\phi$
directions respectively.

We know that every KV is an MC, but the converse is not always
true. As given by Carot et al. [4], if $T_{ab}$ is non-degenerate,
$det(T_{ab})\neq 0$, the Lie algebra of the MCs is finite
dimensional. If $T_{ab}$ is degenerate, i.e., $det(T_{ab})=0$, we
cannot guarantee the finite dimensionality of the MCs. Thus for
the Bianchi types I and III, and Kantowski-Sachs cosmological
models, the Lie algebra of MCs is finite dimensional if $T_{ab}$
is non-degenerate and may be infinite dimensional if $T_{ab}$ is
degenerate. Now $T_{ab}$ is degenerate if any of $T_{00},T_{11}$
and $T_{22}$ vanishes. Since in case (2d) $T_{22}\xi^0=0$, that
Lie algebra of MCs is infinite dimensional. Further, in this case,
it should be noted that we have obtained MCs identical to KVs for
the non-degenerate $T_{ab}$. We observe that the cases (1a),
(2a-c), (3a-c), which contain the component $\xi^0$, have proper
MCs and the cases (2e) and (2f), which do not contain $\xi^0$,
have non-proper MCs. Also, it follows from the case (3d) that for
the Bianchi type I spacetime, some of the MCs are proper but for
the Bianchi type III and Kantowski-Sachs metrics, MCs are
non-proper.

For the Bianchi types II, VIII and IX, spacetimes, we have one MC
in the case (1). Here $\xi^i (i=1,2,3)$ gives a translation along
$x,y$ and $z$ directions respectively. The cases (2a) and (2c)
either give two or one MC according to the constraint while the
cases (2b) and (2f) yield one. In the case (2d), we have either
arbitrary MCs or one according to the constraint whereas in the
case (2e), we get two. Similarly, the case (3b) either give three
or two MCs and the case (3d) either yield $3+\xi^1(y,z)$ or two.
For these cosmological models, it should be noted that the cases
(1a), (2a-c), (3b), types of symmetry vectors are proper MCs while
the cases (1b), (1c), (2d-f) and (3d) give to non-proper MCs.

We remark that RCs obtained by Yavuz et al. [11,12] are similar to
MCs. However, the constraint equations are different. If we solve
these constraint equations, we may have a family of spacetimes. We
have obtained three cosmological models which turn out to perfect
fluid dust solutions. It would be interesting to look for further
solutions from these constraint equations.

\newpage

\begin{description}
\item {\bf Acknowledgments}
\end{description}

The author would like to thank Prof. Chul H. Lee for his
hospitality at the Department of Physics and Korea Scientific and
Engineering Foundation (KOSEF) for postdoc fellowship at Hanyang
University Seoul, KOREA. I am also grateful to Prof. Asghar Qadir
for his useful comments during its write up.

\vspace{2cm}

{\bf \large References}

\begin{description}

\item{[1]} Melfo, A., Nunez, L. Percoco, U. and Villapba, J. Math. Phys. {\bf 33}(1992)2258.

\item{[2]} Bokhari, A.H. and Qadir, A. J. Math. Phys. {\bf
34}(1993)3543; J. Math. Phys. {\bf 31}(1990)1463.

\item{[3]} Bokhari, A.H., Amir, M.J. and Qadir, A. J. Math. Phys. {\bf
35}(1994)3005;\\ Bokhari, A.H., Qadir, A., Ahmad, M.S. and Asghar,
M. J. Math. Phys. {\bf 38}(1997)3639;\\ Qadir, A. and Ziad, M.
Nuovo Cimento B {\bf 113}(1998)773;\\ Bokhari, A.H., Kashif, A.R.
and Qadir, A. J. Math. Phys. {\bf 41}(2000)2167.

\item{[4]} Carot, J., da Costa, J. and Vaz, E.G.L.R. J. Math. Phys. {\bf 35}(1994)4832.

\item{[5]} Carot, J., Numez, L.A. and Percoco, U. Gen. Rel and Grav. {\bf
29}(1997)1223.

\item{[6]} Noether, E. Nachr, Akad. Wiss. Gottingen, II, Math. Phys. Kl{\bf
2}(1918)235.

\item{[7]} Davis, W.R. and Katzin, G.H. Am. J. Phys. {\bf 30}(1962)750.

\item{[8]} Katzin, G.H., Levine, J. and Davis, H.R. J. Math. Phys.{\bf
10}(1969)617.

\item{[9]} Carot, J. and da Costa, J. {\it Proc. of the 6th Canadian Conf. on General Relativity and
Relativistic Astrophysics} Fields Inst. Commun. 15, Amer. Math.
Soc. WC Providence, RI(1997)179.

\item{[10]} Hall, G.S., Roy, I. and Vaz, L.R. Gen. Rel and Grav. {\bf 28}(1996)299.

\item{[11]} Yavuz, $\dot{I}$. and Camci, U. Gen. Rel and Grav. {\bf 28}(1996)691.

\item{[12]} Camci, U., Yavuz, $\dot{I}$. Baysal, H., Tarhan, $\dot{I}$. and Yilmaz, $\dot{I}$.
 Int. J. of Mod Phys. D (to appear 2000).

\item{[13]} MacCallum, M.A.H., {\it In General Relativity: An Einstein Centenary Survey},
eds. Hawking, S. and Israel, W.(Cambridge Univ. Press, 1979)533.

\item{[14]} Baofa, H. Int. J. Theor. Phys. {\bf 30}(1991)1121.

\item{[15]} Lorenz, D. J. Phys. A: Math. Gen.{\bf 15}(1982)2809.

\item{[16]} Banerjee, A.K. and Santos, N.O. Gen. Rel and Grav. {\bf 16}(1984)217.

\item{[17]} Sing, T. and Agrawal, A.K. Astrophysics and Space Science {\bf 191}(1991)61.

\end{description}

\end{document}